\documentstyle[12pt,fleqn,epsfig]{article}

\newcommand{\be}{\begin{equation}}
\newcommand{\ee}{\end{equation}}
\newcommand{\bear}{\be\begin{array}}
\newcommand{\eear}{\end{array}\ee}
\newcommand{\bea}{\begin{eqnarray}}
\newcommand{\eea}{\end{eqnarray}}

\catcode`@=11
\newtoks\@stequation
\def\subequations{\refstepcounter{equation}%
\edef\@savedequation{\the\c@equation}%
 \@stequation=\expandafter{\theequation}
 \edef\@savedtheequation{\the\@stequation}
 \edef\theequation{\theequation}%
 \setcounter{equation}{0}%
 \def\theequation{\theequation\alph{equation}}}
\def\endsubequations{\setcounter{equation}{\@savedequation}%
  \@stequation=\expandafter{\@savedtheequation}%
  \edef\theequation{\the\@stequation}\global\@ignoretrue 
\noindent}
\catcode`@=12

\begin{document}

\begin{flushright}
UCLA/04/TEP/52\\
December 2004
\end{flushright}

\begin{center}
\bigskip\bigskip

{ \large{\bf COSMOLOGY OF ``VISIBLE'' STERILE NEUTRINOS}}\footnote{Talk given at the ``8th Workshop on Non-Perturbative Quantum Chromodynamics",
     June 7-11, 2004, Paris, France; based on
 work done in collaboration
  with S. Palomares-Ruiz and S. Pascoli (see Ref.1)}

\bigskip\bigskip
{ Graciela Gelmini$^1$}

\bigskip\bigskip

${}^{1}${\em Dept. of Physics and Astronomy, UCLA, 405 Hilgard Ave., Los Angeles, CA 90095-1547}

\end{center}



%
%






\begin{abstract}
We point out that in scenarios with a low reheating temperature 
$T_R << 100$ MeV at the end of (the last episode of) inflation or
 entropy production,
the abundance of sterile neutrinos becomes largely independent of
 their coupling  to active neutrinos. Thus, cosmological bounds become
 less stringent than usually assumed, allowing sterile neutrinos to
be ``visible'' in future experiments.
For example, the sterile neutrino required by the LSND result 
does not have any cosmological problem within these scenarios.


\end{abstract}

\section{Why Sterile Neutrinos?}
In the Standard Model of Elementary Particles (SM) there are 
three massless active
neutrinos, $\nu_\alpha$, coupled to the W and Z weak gauge 
bosons through gauge couplings. In
trivial extensions of the SM there are one or more
``sterile'' neutrinos, $\nu_s$, not coupled directly to the Z
and W bosons, coupled to the Higgs boson and the active
 neutrinos through terms in the Lagrangian which
yield non-zero neutrino masses. All extensions of the SM need to be
explored in any event, but in the case of sterile neutrinos
 we have the extra motivation of
knowing that neutrinos have non-zero masses. In fact the 
solar mass-square difference $\Delta m^2_{12} \simeq 8.1 \times 10^{-5}$ eV$^2$
and the atmospheric mass difference
$\Delta m^2_{23} \simeq 2.2 \times 10^{-3}$ eV$^2$~\cite{deltam}
 emply that there are at least three
neutrino mass eigenstates. If confirmed, the LSND result~\cite{lsnd}
 (soon to be tested by MiniBooNE~\cite{miniboone}) can be explained 
with a third square mass difference requiring a fourth neutrino 
mass eigenstate, and, consequently, the existence of at least one sterile 
neutrino. However,  the usual belief is that the existence of
 this sterile  neutrino  is rejected 
by cosmology (see for example Ref.~\cite{cirelli}). In fact,
 all sterile neutrinos which could be found in laboratory 
experiments in the near future, would have problems with usual 
cosmological bounds.

\section{Why ``Visible'' Sterile Neutrinos?}

Here we call ``visible" the sterile neutrinos
which could be found in laboratory experiments. 
In fact, in order to be found, these
sterile neutrinos would necessarily have relatively 
large active-sterile mixings $\sin\theta$.
In the approximation of two-neutrino mixing the interaction neutrino 
eigenstates $|\nu_{\alpha,s}\rangle$ ($\alpha$
 stands for $e$, $\mu$ or $\tau$) are
$|\nu_\alpha \rangle = \cos \theta |\nu_1\rangle +
\sin\theta | \nu_2 \rangle$ and 
{$|\nu_s \rangle = - \sin \theta |\nu_1\rangle
+ \cos \theta | \nu_2 \rangle$,  where
 $|\nu_{1,2}\rangle$ are the neutrino mass
eigenstates. We assume masses 
$m_1 < m_2 \equiv m_s$ (we call $m_s$ the mass
of the heavier, mostly sterile, neutrino mass eigentate).
 
The LSND result, would require the third mass-square difference to be 
$\Delta m^2 \simeq 1 eV^2$, and 
$\sin^2{2 \theta_{LSND}} \sim 0.001$, where,
 for small mixing angles,
$\theta_{LSND} =\theta_{se} \theta_{s\mu}$ (and
 $\sin \theta_{se}$,
$\sin \theta_{s\mu}$ are the two-neutrino mixings 
of $\nu_s$ with $\nu_e$ and with  $\nu_\mu$ respectively).
  But  sterile neutrinos could be
 found in other experiments too: in reactor-$\bar{\nu}_e$ experiments if
 $\Delta m^2 \simeq$  eV$^2$, and
 $\sin^2{2 \theta} \sim 0.1$; in accelerator-$\bar{\nu}_e$ and
$\nu_\mu$ $\bar{\nu}_\mu$ disappearance experiments if
$\Delta m^2 \simeq 10 eV^2$
and $\sin^2{2 \theta} \sim 0.01 -0.001$; in $\beta$--decay experiments
if $\Delta m^2 \simeq$ keV$^2$ and 
$\sin^2{2 \theta} \sim 0.1 -0.001$ and in $(\beta \beta)_{0 \nu}$ --decays
if  $(m_s \sin^{2}{2 \theta}) <$ 4 eV (see Ref.\cite{gpp}
 and references therein). All these
required active-sterile neutrino mixings  much larger that 
standard bounds on dark 
matter abundance allow for. The reason is that
the sterile neutrinos without extra-SM interactions 
considered here,  are produced in the early universe through
 their mixing  with active
neutrinos~\cite{barbieri}  and (see, for example in
Fig.~2 of  Ref.~\cite{abazajian})   have an acceptable abundance only
if their mixing  is  small, for example
$\sin^2 2 \theta < 10^{-5}$ for masses $m_s \leq 10$~eV. This conclusion
follows from standard assumptions made about  the
history of the Universe before nucleosynthesis, which could be different.

\section{Why a Non-Standard Cosmology?}

 Dodelson and Widrow~\cite{dodelson} (see
also Ref.\cite{dolgov}) provided the first analytical calculation of
the production of  sterile neutrinos (without
 extra-SM interactions) in the early Universe, under the
assumption (which we maintain here, for simplicity) 
of a negligible primordial lepton
number, say  as small as the
baryon number in the Universe $L_\nu \simeq 10^{-10}$.

 Because the mass eigenstates $\nu_{1,2}$ evolve with different phases,
 $\approx e^{-i t m_i^2/2E}$
for $E>> m_i$ (we are considering here
neutrinos that are relativistic at production),  an interaction eigenstate
$\nu_\alpha$ produced at $t=0$,
evolves into a mixed state at a later time $t$,
$\nu(t)= a(t)\nu_\alpha+b(t)\nu_s$ . The probability
of $\nu_\alpha$ to become $\nu_s$ at a time $t$ is
\begin{equation}
P(\nu_\alpha \rightarrow \nu_s)= |b(t)|^2= \sin^2 2 \theta
\sin^2 \left(\frac{t}{\ell} \right)~,
\end{equation}
where $\ell=\Delta m^2/ 2E$ is the vacuum oscillation length. 
Matter effects  change $\ell$ into the oscillations length in matter,
 $\ell_m$, and $\sin^2 2 \theta$ into
$\sin^2 2 \theta_m = \left(  \ell_m^2  /  \ell^2  \right)  \sin^2 2 \theta $.
Collisions force the wave function to collapse after a charateristic time
$t_{coll}$. In  the early Universe
 $t_{coll} >> \ell_m$, so the 
$\sin^2\left( t_{coll} / {\ell_m}\right)$ factor appearing in the probability
 $P$ averages to $1/2$ (this is called  the ``averaging regime'').

The rate of production of sterile neutrinos $\Gamma_s$ is given by the
rate of interaction of active neutrinos $\Gamma_\nu$
 multiplied by the probability that
in each collision  the neutrino state would collapse into a sterile
neutrino, i.e.
$\Gamma_s \simeq P(\nu_\alpha \rightarrow \nu_s)\Gamma_\nu $,
 which in the averaging regime is 
 $\Gamma_s \simeq \left( \ell_m^2/ \ell^2 \right)\sin^2 2 \theta~ \Gamma_\nu$.
 With a negligible lepton number,
\begin{equation}
{\ell_m \simeq \frac{\ell}
{\left\{ \sin^2 2\theta + \left[ \cos
2\theta -{\frac{ 2 E~ V^T}{\Delta m^2}}\right]^2\right\}^{1/2}}}
\end{equation}
where $V^T\sim  T^5$ is a  thermal potential
due to finite temperature effects~\cite{raffeltmeffects}
 and $T$ is the temperature of
the Universe.
From this equation we see that at low temperatures  the term containing  $V^T$
 is negligible, thus  matter effects are negligible, so
$\ell_m \simeq \ell$ as in vacuum and  the rate of production of
 sterile neutrinos decreases very fast with decreasing $T$, i.e.
$\Gamma_s \simeq \Gamma_\nu \simeq n \sigma  \sim T^5$ ($n\sim T^3$ is
the number density of particles and $\sigma \sim T^2/M_Z^4$
 is the weak cross section).
 At high enough
 temperatures  instead, the  $V^T$ term dominates, so
$\left( \ell_m / \ell \right) \simeq \Delta m^2 / (V^T 2 E)$, and
 the rate of production of sterile neutrinos increases very 
fast as the temperature decreases, i.e. 
$\Gamma_s \simeq \left(\Delta m^2 / V^T 2 E \right)^2
\Gamma_\nu \sim T^{-7}$.  Thus the rate of production of sterile
 neutrinos has a sharp maximum. The temperature of
 maximum production is~\cite{dodelson} 
 \begin{equation}
 T_{\rm max} \approx 130 {\rm\,
MeV}\left(\frac{m_s}{1{\rm\,keV}}\right)^{1/3}~.
\end{equation}
It is now clear that if  the temperature
of the Universe is always smaller than $ T_{\rm max}$
the production  of sterile neutrinos is suppressed.

\section{Low Reheating Temperature Scenarios}

In inflationary models, or after a late period of entropy production,
 the beginning of the radiation dominated era of
the Universe results from the decay of coherent oscillations of a
scalar field, and the subsequent thermalization of the decay products
into a thermal bath, at the so called ``reheating temperature" $T_R$.
This temperature may have been as low as 0.7 MeV~\cite{kohri} (a
recent analysis strengthens this bound to $\sim$ 4
MeV~\cite{hannestad}). It is well known that a low reheating
temperature inhibits the production of particles which are
non-relativistic or would decouple  at $T > T_R$~\cite{giudice1}. The
final number density of active neutrinos starts departing from the
standard number for $T_R \leq 8$~MeV but stays within 10\% of it for
$T_R \geq 5$~MeV~\cite{giudice2}. 
For $T_R = 1$~MeV the number of
$\nu_{\mu,\tau}$ would be about 2.7\% of the standard number. This would have
allowed one of the active neutrinos to be a warm dark matter (WDM)
candidate (with mass in the keV range), as proposed in Ref. \cite{giudice2},
 if this were not forbidden by
experimental bounds.

Following 
Ref.~\cite{dodelson}, but considering that the production of sterile
neutrinos starts when the temperature of the universe is $T_R <
T_{max}$, the $\nu_s$ distribution function turns out to
be~\cite{gpp}  
\begin{equation}
f_s(E,T)\simeq 3.2~d_{\alpha}\left(\frac{T_{R}}{5~{\rm MeV}}\right)^3 
\!\!\!\! \sin^22\theta
\left(\frac{E}{T}\right) f_\alpha(E,T)
\label{distribution}
\end{equation}
where $d_{\alpha}= 1.13$ for $\nu_{\alpha}= \nu_e$ and $d_{\alpha}=
0.79$ for $\nu_{\alpha}= \nu_{\mu,\tau} $ \cite{fermi}. 

In the calculation the active neutrinos are assumed to have the usual
thermal equilibrium distribution $f_A = (\exp{E/T} + 1)^{-1}$, thus,
following Ref.~\cite{giudice2}, we restrict ourselves to
 $T_R \geq 5 $~MeV. For simplicity, we also restrict
 ourselves to the case
of mass $m_s < 1$~ MeV. 
Moreover,
 for values of $m_s$ such that 
the finite temperature potential~\cite{raffeltmeffects}
 $V^T \ll m_s^2 / 2 E $,
 all matter effects disappear, so the oscillations are as in the vacuum
and no dependence on $m_s$ remains (notice that these oscillations 
are in the averaging regime). 
For $T_{R} = 5$ MeV, the  specific value 
used for the figures, this happens for $m_s \geq 0.2$ eV (0.1 eV)
for $\nu_e \leftrightarrow \nu_s$ ($\nu_{\mu,\tau} \leftrightarrow
\nu_s$). 

The resulting number fraction of sterile over active
neutrinos plus antineutrinos  depends only on
the active-sterile mixing angle and the reheating temperature,
\begin{equation}
\frac{n_{\nu_s}}{n_{\nu_\alpha}} \simeq 10~d_{\alpha}~
\sin^22\theta \left(\frac{T_{R}}{5~{\rm MeV}}\right)^3~.
\end{equation}
Thus,  a low
reheating temperature insures a small sterile number density, even for
very large active-sterile mixing angles. This makes sterile neutrinos in our
scenario potentially detectable in future experiments. Notice that the
$\nu_s$--number density is independent of the mass of the sterile
neutrinos (contrary to the result of Ref.~\cite{dodelson}). Thus, the
mass density of non-relativistic sterile neutrinos, $\Omega_sh^2 =
(m_s \; n_{\nu_s}/\rho_c) h^2$ depends linearly on the mass and on
$\sin^22\theta$, 
\begin{equation}
\Omega_sh^2 \simeq 0.1~d_{\alpha}~
 \left(\frac{\sin^22\theta}{10^{-3}}\right)            
 \left(\frac{m_s}{1~\rm{keV}}\right) 
\left(\frac{T_R}{5~\rm{MeV}}\right)^3~.
\end{equation}
The condition $\Omega_sh^2 \leq \Omega_{DM} h^2 = 0.135$~\cite{WMAP}
rejects the triangular dark gray region of masses and mixings shown in 
 Figs.~1  and 2. The values of masses and mixings for which sterile
 neutrinos constitute 10\% of the dark matter are also shown with a
 dotted line.

\begin{figure}
\centerline{\epsfxsize=3.4in \epsfbox{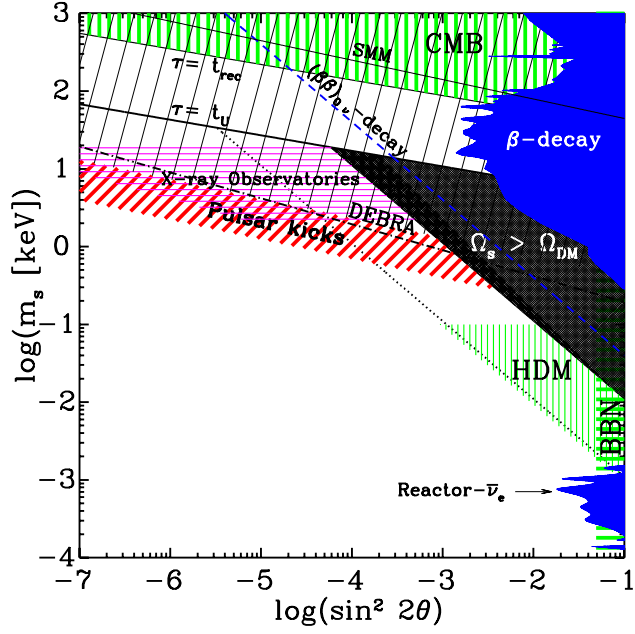}}
\caption{\label{nue}
Bounds and sensitivity regions for $\nu_e \leftrightarrow \nu_s$
oscillations for $T_R \sim 5$~MeV.}
\end{figure}
\begin{figure}
\centerline{\epsfxsize=3.4in \epsfbox{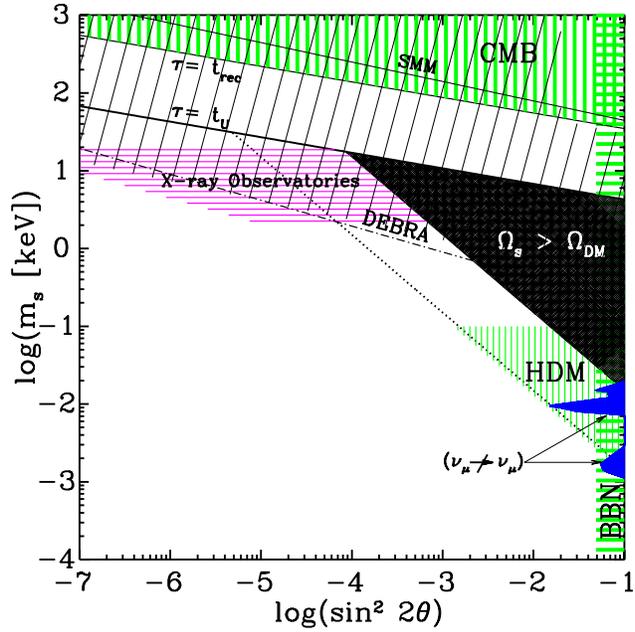}}
\caption{\label{numu} Same as Fig.~1 for $\nu_{\mu, \tau}
  \leftrightarrow \nu_s$. For $\nu_\tau  \leftrightarrow \nu_s$ the
 darkest gray-blue excluded region does not apply.}
\end{figure}

Figs.~1 and 2  show  bounds for $\nu_{\alpha}= \nu_e$ and
$\nu_{\alpha}= \nu_{\mu,\tau}$, respectively, and for $T_R =5$~MeV. 
They show that $\nu_s$ in our scenario could only be 
part of the hot dark matter (HDM) in the Universe
(i.e. $m_s <<$ keV), while  neutrinos with $ m_{s} > 1$~keV 
are disfavored, if not rejected, as WDM ($m_s\simeq $ keV) 
or CDM (cold dark matter, $m_s>>$ keV), by bounds coming from
 supernovae cooling (which exclude the region diagonally 
hatched  with thin lines) and
astrophysical bounds due to radiative decays (which exclude 
the region above  the line labeled ``DEBRA" for ``diffused extragalactic
background radiation''), explained in detail in
Ref. 1. The region denoted with ``X-ray Observatories" 
could be rejected by observations of 
galaxy clusters by the Chandra observatory\cite{abazajian2}, 
and the region labeled ``Pulsar
kicks" shows where sterile neutrinos could possibly explain the large velocity
of pulsars~\cite{FKMP}. The SN1987A bound on
neutrino radiative decays, excludes the region above the
line labeled ``SMM" (for the ``Solar Maximum Mission''
 satellite) in both figures.
The 3$\sigma$ upper bound imposed by big bang nucleosynthesis on
 any extra contribution to the energy density $\Delta N_\nu\leq 0.73$ (see
 Fig.~7 of Ref.~\cite{steigman}), translates into the
 vertical excluded band labeled ``BBN"
  in Figs.~1 and 2.

Experimental bounds are also shown (see Ref. \cite{gpp} for
more details). Negative results from reactor-$\bar{\nu}_e$ and  
accelerator-$\nu_\mu$ disappearance experiments reject
 the darkest/blue regions so labeled in Fig.1 and in Fig. 2
 respectively. The absence of
kinks in the  $e^-$  spectra in $\beta$--decays constrain
 the $\nu_e - \nu_s$
mixing (darkest/blue region in Fig.~1).
If neutrinos are Majorana particles, present bounds on 
 neutrinoless double beta decay
constrain the contribution of the mostly sterile 
neutrino to the effective $\nu_e$ Majorana mass,
which conservatively translates into  the upper bound
 $\left( m_s  \ \sin^2{2 \theta} \right) < 4 ~ {\rm eV} $
 (allowed region below the dashed line in Fig.~1).

In conclusion, in cosmological  scenarios with  a low 
reheating temperature at
the end of the last episode of  inflation or entropy
 production, such as $T_R \sim 5$~MeV,
  the coupling of sterile to active neutrinos
 can be much larger than standard cosmological scenarios  permit. For
example, the sterile neutrino required by the LSND result does not
have any cosmological  problem. In these scenarios,    the  baryon
 asymmetry and the bulk of the dark matter  in the Universe
  should also originate in novel ways.

The experimental discovery of a sterile neutrino in the region of
$m_s-\sin^2 {2\theta}$ opened up in this paper, would require an
unusual cosmology, such as one with a low reheating temperature as
presented here.

\section*{Acknowledgments}

This work was
supported in part by the DOE grant DE-FG03-91ER40662, Task C and NASA
grant NAG5-13399.


\begin{thebibliography}{0}


\bibitem{gpp} G.~Gelmini, S.~Palomares-Ruiz and S.~Pascoli,
Phys.\ Rev.\ Lett.\  {\bf 93}, 081302 (2004).

\bibitem{deltam} M.~Maltoni {\it et al.}
hep-ph/0405172.


\bibitem{lsnd} A.~Aguilar {\it et al.}, Phys.\ Rev.\ D {\bf 64}, 112007 (2001).

\bibitem{miniboone} E.~Church {\it et al.}, FERMILAB-P-0898;
I.~Stancu {\it et al.},  http://www-boone.fnal.gov/.

\bibitem{cirelli} M. Cirelli {\it et al.}, hep-ph/0403158.

\bibitem{barbieri} R.~Barbieri and A.~Dolgov, Phys. Lett. B{\bf 237},
440 (1990) and
Nucl. Phys. B{\bf 349}, 742 (1991); K.~Kainulainen,
Phys.\ Lett.\ B {\bf 244}, 191 (1990);
K.~Enqvist, K.~Kainulainen and J.~Maalampi,
Phys. Lett. B{\bf244}, 186 (1990) and Phys. Lett. B{\bf 249}, 531 (1990).


\bibitem{abazajian} K.~Abazajian, G.~M.~Fuller and M.~Patel,
Phys.\ Rev.\ D {\bf 64}, 023501 (2001).

    
    
\bibitem{dodelson} S.~Dodelson and L.~M.~Widrow,  Phys. Rev. Lett. 
{\bf 72}, 17 (1994).

\bibitem{dolgov} A.~D.~Dolgov and S.~H.~Hansen,
Astropart.\ Phys.\  {\bf 16}, 339 (2002).


\bibitem{raffeltmeffects}   
D.~Notzold and G.~Raffelt,
Nucl.\ Phys.\ B {\bf 307}, 924 (1988).
\bibitem{kohri}
M.~Kawasaki, K.~Kohri and N.~Sugiyama,
Phys.\ Rev.\ Lett.\  {\bf 82}, 4168 (1999)
and Phys.\ Rev.\ D {\bf 62}, 023506 (2000).

\bibitem{hannestad}
S.~Hannestad,
Phys.\ Rev.\ D {\bf 70}, 043506 (2004).

\bibitem{giudice1} 
G.~F.~Giudice, E.~W.~Kolb and A.~Riotto,
Phys.\ Rev.\ D {\bf 64}, 023508 (2001).


\bibitem{giudice2} G.~F.~Giudice {\sl et al.}, 
Phys.\ Rev.\ D {\bf 64}, 043512 (2001).


\bibitem{fermi}
A.~D.~Dolgov {\sl et al.}, 
Astropart.\ Phys.\  {\bf 14}, 79 (2000).


\bibitem{WMAP}
D.~N.~Spergel {\it et al.},
Astrophys.\ J.\ Suppl.\  {\bf 148}, 175 (2003).

\bibitem{abazajian2} K.~Abazajian, G.~M.~Fuller and W.~H.~Tucker,
Astrophys.\ J.\  {\bf 562}, 593 (2001).

\bibitem{FKMP}
G.~M.~Fuller {\sl et al.}, 
Phys.\ Rev.\ D {\bf 68}, 103002 (2003).

\bibitem{steigman} 
V.~Barger {\sl et al.},
Phys.\ Lett.\ B {\bf 566}, 8 (2003).


\end{thebibliography}
\end{document}